\documentclass[showpacs,prb,twocolumn]{revtex4}%
\usepackage{amsfonts}
\usepackage{amsmath}
\usepackage{amssymb}
\usepackage{graphicx}%
\ifx\pdfoutput\relax\let\pdfoutput=\undefined\fi
\newcount\msipdfoutput
\ifx\pdfoutput\undefined\else
\ifcase\pdfoutput\else
\ifx\paperwidth\undefined\else
\ifdim\paperheight=0pt\relax\else\pdfpageheight\paperheight\fi
\ifdim\paperwidth=0pt\relax\else\pdfpagewidth\paperwidth\fi
\fi\fi\fi
\begin{document}
\title{Spin relaxation in diluted magnetic semiconductor quantum dots}
\author{W. Yang}
\author{K. Chang}
\altaffiliation[To whom correspondences should be sent to: ]{kchang@red.semi.ac.cn}

\affiliation{NLSM, Institute of Semiconductors, Chinese Academy of Sciences, P. O. Box 912,
Beijing 100083, China}


\begin{abstract}
Electron spin relaxation induced by phonon-mediated \textit{s-d }exchange
interaction in a II-VI diluted magnetic semiconductor quantum dot is
investigated theoretically. The electron-acoustic phonon interaction due to
piezoelectric coupling and deformation potential is included. The resulting
spin lifetime is typically on the order of microseconds. The effectiveness of
the phonon-mediated spin-flip mechanism increases with increasing Mn
concentration, electron spin splitting, vertical confining strength and
lateral diameter, while it shows non-monotonic dependence on the magnetic
field and temperature. An interesting finding is that the spin relaxation in a
small quantum dot is suppressed for strong magnetic field and low Mn
concentration at low temperature.

\end{abstract}
\maketitle

\section{Introduction}

The spin of the electron in low-dimensional semiconductor structures has
received intense interest in recent years due to its potential applications in
spintronic devices and quantum information processing
technologies.\cite{Wolf,Loss} To improve the performance of such devices, the
decoherence of the electron spin due to coupling to environmental degrees of
freedom should be minimized.
Theoretical\cite{Nazarov,Nazarov2,Woods,Efros,Semenov} and experimental
investigations\cite{Awschalom,Paillard} have shown that the electron spin
could have a extremely long relaxation time in nonmagnetic semiconductor
quantum dots (QD's), compared with that in the bulk or quantum wells.
Theoretical works proposed that the diluted magnetic semiconductor (DMS) QD's
can be used as spin aligners, spin memories as well as spin
qubits.\cite{Loss2} DMS QD's offer us a new flexibility in manipulating
carrier spins, since the spin properties can be strongly influenced by
applying an external magnetic field or varying the temperature.\cite{KChang}
Various relaxation mechanisms of the electron spin come from different
coupling to the environment, i.e., magnetic impurity, nuclear spin, and
spin-orbit interaction. It is important to identify the dominant mechanism of
spin relaxation for a particular system. Previous
theoretical\cite{Bastard,Egues} and experimental\cite{Awschalom2,Akimoto}
works on spin relaxation in DMS quantum wells have indicated that the
\textit{s-d }exchange interaction between band electrons and localized spins
of magnetic ions is the dominant spin-flip mechanism, leading to electron spin
lifetime of the order of picoseconds. However, to the best of our knowledge,
there is no study on the spin relaxation induced by the \textit{s-d }exchange
interaction in DMS QD's. Detailed theoretical and experimental investigations
are necessary to gain physical insight into the spin relaxation process in
such structures.

In this paper, we investigate theoretically the spin relaxation of the lowest
Zeeman doublet in vertical II-VI DMS QD's. The spin-flip scattering caused by
the acoustic phonon-mediated \textit{s-d} exchange interaction between the
conduction electron and Mn ions is considered. The electron-acoustic phonon
coupling includes the piezoelectric and deformation potential interactions.
Since the first-order spin-flip process through direct scattering by the Mn
ions is generally blocked by the energy-matching condition, we consider the
second-order process involving the emission or absorption of a phonon. Our
calculation shows that this phonon-mediated spin-flip scattering leads to
electron spin lifetime typically of the order of microseconds. The
effectiveness of this mechanism increases significantly with increasing Mn
concentration, electron spin splitting, vertical confining strength and
lateral diameter, while it shows non-monotonic dependence on the magnetic
field and temperature, due to the competing effect of the electron spin
splitting, the phonon number and the correlation function of the Mn ions. It
is interesting to notice that the spin relaxation of electrons in the lowest
Zeeman doublet is suppressed in a strong magnetic field at low temperature for
II-VI DMS QD's with low Mn concentration.

The rest of this paper is organized as follows: the theoretical model and
formula of the spin-flip scattering rate (SFR) induced by phonon-mediated
\textit{s-d} exchange interaction are derived in sec. II. Numerical results
and discussions for the SFR as a function of magnetic field, as well as its
dependence on the Mn concentration, QD size and temperature are given in sec.
III and we give a brief conclusion in sec. IV.

\section{Theory}

We consider II-VI DMS QD's subjected to a perpendicular magnetic field.
Assuming an infinite deep well along the growth direction (the $z$ axis) and a
in-plane parabolic confining potential, the electron wave function can be
written as $\psi(\mathbf{r})=\chi(z)\phi(\rho,\varphi)$, where $\chi
(z)=\sqrt{2/z_{0}}\sin(\pi z/z_{0})$ is the ground state wave function along
the $z$ axis (we have assumed that the vertical confinement is strong and only
the lowest energy level is relevant), $z_{0}$ is the width of the well, and
$\phi(\rho,\varphi)$ is the in-plane wave function determined by the
two-dimensional Hamiltonian $H=H_{0}+H_{s-d}+H_{e-p}$. The first term
\begin{equation}
H_{0}=\frac{(\mathbf{p}+e\mathbf{A})^{2}}{2m^{\ast}}+\frac{1}{2}m^{\ast}%
\omega_{0}^{2}\rho^{2}+\frac{1}{2}g^{\ast}\mu_{B}B\sigma_{z} \label{H0}%
\end{equation}
is the electron Hamiltonian in the external magnetic field and parabolic
potential.\ Here $m^{\ast}$ is the electron effective mass, $\mathbf{A}%
=(-By/2,Bx/2,0)$ is the vector potential, $\omega_{0}$ characterizes the
lateral confinement strength, $g^{\ast}$ is the intrinsic electron g-factor,
and $\sigma_{z}$ is the $z$-component of the Pauli matrices. The second term%
\begin{equation}
H_{s-d}=-%
{\displaystyle\sum\limits_{i}}
J(\mathbf{r}-\mathbf{R}_{i})\mathbf{s}\cdot\mathbf{S}_{i} \label{Hsd}%
\end{equation}
describes the \textit{s-d} exchange interaction between the electron
($\mathbf{s}$) and the localized Mn ion ($\mathbf{S}_{i}$), where
$J(\mathbf{r})$ is the \textit{s-d} coupling integral, and the summation runs
over all the Mn sites. The last term$\cite{Nazarov2}$
\begin{equation}
H_{e-p}=%
{\displaystyle\sum\limits_{\mathbf{q},\nu}}
\alpha_{\nu}(\mathbf{q})(b_{\mathbf{q},\nu}e^{i\mathbf{q}\cdot\mathbf{r}%
}+b_{\mathbf{q},\nu}^{+}e^{-i\mathbf{q}\cdot\mathbf{r}}) \label{Hep}%
\end{equation}
describes the interaction between the electron and acoustic phonon, where
$b_{\mathbf{q}\nu}(b_{\mathbf{q}\nu}^{+})$ is the annihilation (creation)
operator of the bulk phonon mode with wave vector $\mathbf{q}$ and branch
$\nu$.

The \textit{s-d} exchange term is divided into a mean-field part and a
fluctuating part, $H_{s-d}=H_{s-d}^{0}+V_{s-d},$ where $H_{s-d}^{0}=\sigma
_{z}\Delta_{sd}/2,$ $V_{s-d}=-\sum\nolimits_{i}J(\mathbf{r}-\mathbf{R}%
_{i})(s^{(+)}S_{i}^{(-)}+s^{(-)}S_{i}^{(+)})/2,$ $s^{(\pm)}=s_{x}\pm is_{y},$
$S_{i}^{(\pm)}=S_{i}^{x}\pm S_{i}^{y},$ $\Delta_{sd}=-N_{0}\alpha
x\left\langle S_{z}\right\rangle $ is the exchange splitting$,$ $N_{0}$ is the
number of unit cells per unit volume, $\alpha=\left\langle \phi_{c}\left\vert
J(\mathbf{r})\right\vert \phi_{c}\right\rangle /\Omega$ ($\Omega$ is the unit
cell volume, $\phi_{c}$ is the Bloch function at the bottom of the conduction
band) is the \textit{s-d }exchange coupling constant, $x$ is the fractional
occupation factor of the cation sites by the Mn ions,
\begin{equation}
\left\langle S_{z}\right\rangle =-S_{0}B_{S}\left[  \frac{g_{Mn}\mu_{B}%
BS}{k_{B}(T+T_{0})}\right]  \label{Sz}%
\end{equation}
is the thermal average of the Mn spin, with $S=5/2$ the Mn 3d$^{\text{5}}$
spin, $B_{S}(x)$ the Brillouin function, and $S_{0},$ $T_{0}$ phenomenological
parameters accounting for the antiferromagnetic superexchange between
neighboring Mn ions. Now the total Hamiltonian is divided into two parts,
$H=\bar{H}_{0}+V,$ where%

\begin{equation}
\bar{H}_{0}=\dfrac{p^{2}}{2m^{\ast}}+\dfrac{1}{2}m^{\ast}\omega^{2}\rho
^{2}+\dfrac{1}{2}\omega_{c}L_{z}+\dfrac{1}{2}\sigma_{z}\Delta_{z},
\label{RenormalizedH0}%
\end{equation}
$V=V_{s-d}+H_{e-p}$. Here $\omega=\sqrt{\omega_{0}^{2}+\omega_{c}^{2}/4},$
$\omega_{c}=eB/m^{\ast}$ is the cyclotron frequency$,$ $L_{z}$ is the
$z$-component of the orbital angular momentum, $\Delta_{z}=g^{\ast}\mu
_{B}B+\Delta_{sd}$ is the total Zeeman splitting of the electron.

In order to obtain the SFR induced by the \textit{s-d }exchange and
electron-acoustic phonon interaction, we consider the whole system (including
the electron, Mn ions and the phonon bath) transits from an initial state
$\left\vert i\right\rangle =\left\vert l\sigma;M;N\right\rangle $ to all
possible final states $\left\vert f\right\rangle =\left\vert l^{\prime}%
\bar{\sigma};M^{\prime};N^{\prime}\right\rangle $ in which the electron spin
is reversed. Here $\left\vert l\sigma\right\rangle $ is the electron
eigenstate ($l$ stands for the orbital quantum number $(n,m)$, see the
Appendix for details), $\sigma=\pm$ ($\bar{\sigma}=\mp$) denote spin-up
(spin-down) and spin-down (spin-up) state, respectively. $\left\vert
M\right\rangle =\left\vert M_{1z},M_{2z},\cdots\right\rangle $ is the
eigenstate of the Mn ions and $\left\vert N\right\rangle =\prod_{\mathbf{q}%
\nu}\left\vert n_{\mathbf{q}\nu}\right\rangle $ denotes the phonon state. The
transition rate is averaged over the random positions and the initial states
of the Mn ions, as well as the initial states of the phonon system to give the
SFR of the electron from the initial state $\left\vert l\sigma\right\rangle $
to the final state $\left\vert l^{\prime}\bar{\sigma}\right\rangle $, denoted
as $W_{l^{\prime}\bar{\sigma},l\sigma}$.

Since the spin-flip process of electron is always accompanied by the flip of a
Mn spin due to the conservation of the total angular momentum (see Eq.
(\ref{Hsd})), we introduce the renormalized electron energy $E_{l\pm
}=\varepsilon_{l}\pm\Delta_{0}/2$, where $\varepsilon_{l}$ is the orbital
eigenenergy of $\bar{H}_{0}$ (see the Appendix for details) and $\Delta
_{0}=\Delta_{sd}-\Delta_{i}$ is the (renormalized) electron spin splitting,
with $\Delta_{i}=(g_{Mn}-g^{\ast})\mu_{B}B$. Based on second-order
perturbation theory, the transition amplitude between $\left\vert
i\right\rangle $ and $\left\vert f\right\rangle $ is given by
\begin{widetext}%
\begin{align}
T_{fi} &  =%
{\displaystyle\sum\limits_{l_{1}}}
\left[  \frac{\left\langle l^{\prime}\bar{\sigma};M^{\prime};N^{\prime
}|V_{s-d}|l_{1}\sigma;M;N^{\prime}\right\rangle \left\langle l_{1}%
\sigma;M;N^{\prime}|H_{e-p}|l\sigma;M;N\right\rangle }{E_{l^{\prime}%
\bar{\sigma}}-E_{l_{1}\sigma}}\right.  \label{Tfi}\\
&  \left.  +\frac{\left\langle l^{\prime}\bar{\sigma};M^{\prime};N^{\prime
}|H_{e-p}|l_{1}\bar{\sigma};M^{\prime};N\right\rangle \left\langle l_{1}%
\bar{\sigma};M^{\prime};N|V_{s-d}|l\sigma;M;N\right\rangle }{E_{l\sigma
}-E_{l^{\prime}\bar{\sigma}}}\right]  .\nonumber
\end{align}
\end{widetext}
In the first term, the electron first hops from the initial state $\left\vert
l\sigma\right\rangle $ to an virtual state with the same spin $\left\vert
l_{1}\sigma\right\rangle $ through the interaction with a phonon, then it
makes a spin-flip transition to the final state $\left\vert l^{\prime}%
\bar{\sigma}\right\rangle $ through the \textit{s-d} exchange interaction with
one Mn ion. The second term describes the process that the electron is first
scattered to an opposite-spin virtual state $\left\vert l_{1}\bar{\sigma
}\right\rangle $ through the \textit{s-d }exchange interaction with one Mn
ion, then it transits to the final state $\left\vert l^{\prime}\bar{\sigma
}\right\rangle $ via the assistance of a phonon.

The scattering rate of the whole system is obtained from the Fermi golden rule
$W_{fi}=(2\pi/\hbar)\left\vert T_{fi}\right\vert ^{2}\delta(E_{f}-E_{i})$, and
the SFR for the electron system is given by%
\begin{align}
W_{l^{\prime}-,l+}  &  =\frac{1}{4}x(N_{0}\alpha)^{2}G^{-+}\left[
n(\left\vert \Delta_{ll^{\prime}}\right\vert )+\frac{1+\text{sign}%
(\Delta_{ll^{\prime}})}{2}\right]  K_{ll^{\prime}},\label{Wsf2_UpDown}\\
W_{l+,l^{\prime}-}  &  =\frac{1}{4}x(N_{0}\alpha)^{2}G^{+-}\left[
n(\left\vert \Delta_{ll^{\prime}}\right\vert )+\frac{1-\text{sign}%
(\Delta_{ll^{\prime}})}{2}\right]  K_{ll^{\prime}}, \label{Wsf2_DownUp}%
\end{align}
where $G^{-+}=\left\langle S^{(-)}S^{(+)}\right\rangle ,$ $G^{+-}=\left\langle
S^{(+)}S^{(-)}\right\rangle $ are correlation functions of the Mn ions$,$
$S^{(\pm)}=S_{x}\pm iS_{y}$, $\Delta_{ll^{\prime}}=\varepsilon_{l}%
-\varepsilon_{l^{\prime}}+\Delta_{0}$ is the electron energy detuning$,$
$n(E)=\left[  \exp(-E/(k_{B}T))-1\right]  ^{-1}$ is the phonon distribution
function, sign$(x)=1$ for $x>0$ and $-1$ for $x<0$. The kernel $K_{ll^{\prime
}}$\ is given by \begin{widetext}
\begin{equation}
K_{ll^{\prime}}=%
{\displaystyle\sum\limits_{l_{1},l_{2}}}
\left[  \dfrac{S_{l^{\prime}l_{1}l^{\prime}l_{2}}\Gamma_{ll_{1}ll_{2}}}%
{\Delta_{l_{1}l^{\prime}}\Delta_{l_{2}l^{\prime}}}+\dfrac{S_{ll_{1}ll_{2}%
}\Gamma_{l^{\prime}l_{1}l^{\prime}l_{2}}}{\Delta_{ll_{1}}\Delta_{ll_{2}}%
}-\dfrac{2\operatorname{Re}(S_{l_{1}l_{2}ll^{\prime}}\Gamma_{l_{1}ll^{\prime
}l_{2}})}{\Delta_{l_{1}l^{\prime}}\Delta_{ll_{2}}}\right]  , \label{Kernel}%
\end{equation}
\end{widetext}
where
\begin{equation}
S_{l_{1}l_{2}l_{3}l_{4}}=\Omega\int d^{3}\mathbf{R\ }\left\langle
l_{1}|\mathbf{R}\right\rangle \left\langle l_{2}|\mathbf{R}\right\rangle
\left\langle \mathbf{R}|l_{3}\right\rangle \left\langle \mathbf{R}%
|l_{4}\right\rangle \label{OverlapIntegral1}%
\end{equation}
\ is the dimensionless overlap integral, and\begin{widetext}
\begin{equation}
\Gamma_{l_{1}l_{2}l_{3}l_{4}}=\frac{2\pi}{\hbar}\sum\limits_{q,\nu}\left\vert
\alpha_{\nu}(\mathbf{q})\right\vert ^{2}\left\langle l_{1}|e^{i\mathbf{q}%
\cdot\mathbf{r}}|l_{2}\right\rangle \left\langle l_{4}|e^{-i\mathbf{q}%
\cdot\mathbf{r}}|l_{3}\right\rangle \delta(\hbar\omega_{\mathbf{q}\nu
}-\left\vert \Delta_{ll^{\prime}}\right\vert ) \label{PhononIntegral1}%
\end{equation}
\end{widetext}
is related to the spin-conserved phonon-induced transition rate. The explicit
expressions for $S_{l_{1}l_{2}l_{3}l_{4}}$ and $\Gamma_{l_{1}l_{2}l_{3}l_{4}}$
are given in the Appendix.

The spin lifetime $\tau_{l\sigma}$ of a given energy level $\left\vert
l\sigma\right\rangle $ is given by
\begin{equation}
\frac{1}{\tau_{l\sigma}}=\sum\limits_{l^{\prime}}W_{l^{\prime}\bar{\sigma
},l\sigma}, \label{Lifetime}%
\end{equation}
i.e., the sum of the spin-flip scattering rates from $\left\vert
l\sigma\right\rangle $ to all opposite-spin final states $\left\vert
l^{\prime}\bar{\sigma}\right\rangle $.

\section{Numerical results and discussions}

From Eq. (\ref{Kernel}), we notice that the kernel $K_{ll^{\prime}}$ and the
SFR $1/\tau_{l\sigma}$ (see Eq. (\ref{Lifetime})) diverges when the energy of
the intermediate state coincides with the initial or final state. To remove
this divergence, we take into account the finite lifetime of the intermediate
level and add a small broadening parameter (an order-of-magnitude estimate is
0.1meV\cite{Brunner,Gammon,Garcia}) to the energy of the intermediate state to
convert this divergence into a resonance near the degeneracy
point.\cite{Cardona} This broadening parameter is not crucial for our
calculation and would not change the qualitative behavior of the SFR.

We consider Cd$_{1-x}$Mn$_{x}$Te QD's and use the following parameters in our
numerical calculations: $m^{\ast}$=0.096m$_{0}$ (m$_{0}$ is the free electron
mass), $g^{\ast}$=$-$1.6, CdTe mass density $\rho$=5.86 g/cm$^{3}$, lattice
constant $a$=0.6481 nm, $g_{Mn}$=2, $S$=5/2, $N_{0}\alpha$=220 meV, $h_{14}%
$=0.394$\times$10$^{9}$ V/m, sound velocity $C_{l}$=$3083$ m/s, $C_{t}$=1847
m/s. The lateral confining strength of the QD is characterized by the lateral
diameter $d=2\sqrt{\hbar/(m^{\ast}\omega_{0})}$. The dependence of $S_{0},$
$T_{0}$ on the Mn concentration $x$ is obtained from Ref. 20.

Considering the electron occupies the lowest spin-up and spin-down levels
$(n=0,m=0,\pm)$, i.e., the lowest Zeeman doublet in the DMS QD, the SFR's can
be calculated for the spin-up and spin-down states, which are denoted by
$1/\tau_{+}$ and $1/\tau_{-}$, respectively.%

\begin{figure}
[ptb]
\begin{center}
\includegraphics[
height=6.4614cm,
width=8.3088cm
]%
{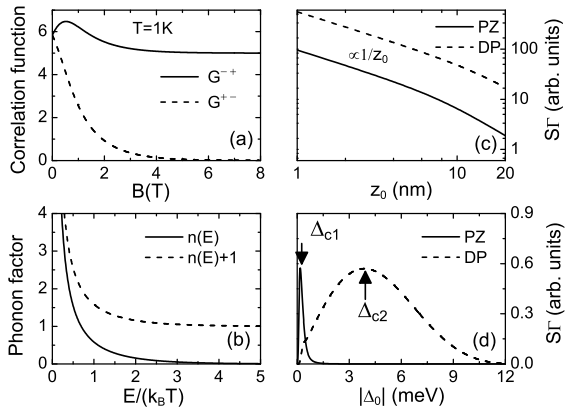}%
\caption{(a) Magnetic field dependence of the Mn correlator $G^{-+}$ (solid
curve) and $G^{+-}$ (dashed curve) at $T$=1 K. (b) Phonon absorption (solid
line) and emission (dashed line) factor. (c) and (d): Piezoelectric coupling
(solid curves) and deformation potential (dashed curves) contribution to the
product $S\Gamma$ as a function of $z_{0}$ and $\left\vert \Delta
_{0}\right\vert $, respectively.}%
\end{center}
\end{figure}
Since the spin-flip transitions to excited orbital levels are energetically
unfavorable, the SFR of the lowest Zeeman doublet is dominated by the
transition between the doublet, i.e., $W_{00+,00-}$ and $W_{00-,00+}$, such
that the electron spin splitting $\Delta_{0}=\Delta_{00,00}$ and the kernel
$K_{00,00}$ are important quantities for $1/\tau_{\pm}$ (see Eq.
(\ref{Wsf2_UpDown}) and Eq. (\ref{Wsf2_DownUp})). From Eq. (\ref{Kernel}), we
note the contribution to $K_{00,00}$ comes mainly from the intermediate level
whose energy is the closest to the initial (or final) state (i.e., the orbital
state $(0,-1)$ in most cases, see the Appendix for details). That is, the term
containing $S_{00,0-1,00,0-1}\Gamma_{00,0-1,00,0-1}$ and $S_{0-1,0-1,00,00}%
\Gamma_{00,0-1,0-1,00}$ (from the Appendix, we see they are equal to each
other, so both terms are denoted as $S\Gamma$ for short) in Eq. (\ref{Kernel})
is the dominant contribution to $K_{00,00}$. Additional contributions to the
SFR $1/\tau_{\pm}$ are the correlation function $G^{-+}$, $G^{+-}$, the phonon
emission factor $n(\left\vert \Delta_{0}\right\vert )+1$ or absorption factor
$n(\left\vert \Delta_{0}\right\vert )$. Therefore, in Fig. 1(a), (b), (c),
(d), we plot schematically the correlation function $G^{-+}$ and $G^{+-}$, the
phonon emission (absorption) factor and the product $S\Gamma$ as a function of
magnetic field, phonon energy, $z_{0}\ $and $\left\vert \Delta_{0}\right\vert
,$ respectively. In Fig. 1(a), we see that $G^{+-}=S(S+1)-\left\langle
S_{z}^{2}\right\rangle -\left\vert \left\langle S_{z}\right\rangle \right\vert
$ decreases monotonically to zero while $G^{-+}=S(S+1)-\left\langle S_{z}%
^{2}\right\rangle +\left\vert \left\langle S_{z}\right\rangle \right\vert $
shows a peak and approaches a constant value with increasing magnetic field.
Physically, this is because $G^{+-}$ ($G^{-+}$) is related to the transition
of an electron from a spin-down (spin-up) initial state to a spin-up
(spin-down) final state (see Eq. (\ref{Wsf2_UpDown}) and Eq.
(\ref{Wsf2_DownUp})). Due to the conservation of the total angular momentum,
the $z$-component of the Mn spin $\left\langle S_{z}\right\rangle $ should
decrease (increase) by one in this process. In a strong magnetic field,
however, all the Mn spins are polarized antiparallel the magnetic field (i.e.,
$\left\langle S_{z}\right\rangle \approx-5/2$), thus the correlation function
$G^{+-}$ tends to vanish and the spin-flip process of the spin-down state is
suppressed. The decrease of $S\Gamma$ with increasing $z_{0}$ (approximately
$S\Gamma\propto1/z_{0}$) and the peak behaviors of $S\Gamma$ as a function of
$\left\vert \Delta_{0}\right\vert $ can be appreciated from Eq.
(\ref{OverlapIntegral}), (\ref{PZPhonon}) and (\ref{DPPhonon}) in the
Appendix. Note in the region where $\left\vert \Delta_{0}\right\vert $ is
small, the dependence of $S\Gamma$ on $\left\vert \Delta_{0}\right\vert $ is
in agreement with Ref. 4. A peculiar feature is $S\Gamma$ vanishes when the
electron spin splitting $\Delta_{0}$ approaches zero, due to the vanishing
energy and, as a result, the vanishing density of states of the involved phonon.

Next, we shall investigate $1/\tau_{\pm}$ as a function of magnetic field for
different temperatures, Mn concentrations and lateral diameters. The effect of
vertical confining length $z_{0},$ lateral diameter $d$ and temperature on
$1/\tau_{\pm}$ is also presented.

\subsection{Strong lateral confinement}

First we consider a small DMS QD with strong vertical confinement $z_{0}$=2 nm
and small lateral diameter $d$=16 nm, such that the vertical and in-plane
orbital energy separations are $\sim$3 eV and $\sim$12 meV$,$ respectively.
The large vertical orbital energy separation ensures only the lowest bound
state is relevant to the spin relaxation, while the in-plane orbital energy
separation which is much larger than $\Delta_{i}$ ensures that the spin-flip
transitions between the lowest Zeeman doublet usually dominate the spin
relaxation process. However, if the Mn concentration is fairly high and the
temperature is sufficiently low, the exchange splitting $\Delta_{sd}$ may
eventually become comparable with the orbital energy separation, then the
lowest spin-up level may cross spin-down excited levels, opening up new spin
relaxation channels for the spin-up state.

\subsubsection{Low Mn concentration}%

\begin{figure}
[ptb]
\begin{center}
\includegraphics[
height=6.2902cm,
width=8.3088cm
]%
{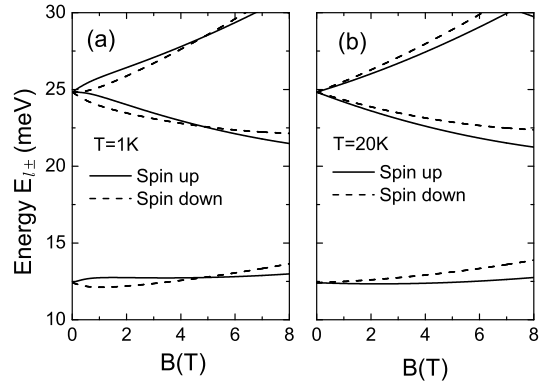}%
\caption{The renormalized spin-dependent electron energy spectra as a function
of magnetic field at (a) $T$=1 K and (b) $T$=20 K, with fixed $z_{0}$=2 nm,
$d$=16 nm, $x$=0.002. The spin-up (spin-down) levels are denoted by solid
(dashed) lines.}%
\end{center}
\end{figure}
The Mn concentration is taken as $x$=0.002$,$ i.e., we take the saturated
exchange splitting $(\Delta_{sd})_{sat}$ $\sim$1 meV. The renormalized
spin-dependent energy spectra for the electron at $T$=1 K and $T$=20 K are
shown in Fig. 2. At low temperature $T$=1 K (Fig. 2(a)), the thermal-averaged
Mn spin $\left\vert \left\langle S_{z}\right\rangle \right\vert $ grows
rapidly with increasing magnetic field (cf. Eq. (\ref{Sz})). As a result, the
exchange splitting $\Delta_{sd}$ increases rapidly to its maximum ($\sim$1
meV) and saturates, while $\Delta_{i}\propto B$ increases smoothly, such that
the electron spin splitting $\Delta_{0}=\Delta_{sd}-\Delta_{i}$ reaches its
maximum at a critical magnetic field, decreases when the magnetic field grows
stronger, and eventually changes its sign at a strong enough magnetic field.
In the high temperature case (see Fig. 2(b)), $\left\vert \left\langle
S_{z}\right\rangle \right\vert $ increases very slowly and $\Delta_{i}$ always
dominates, leading to a negative $\Delta_{0}$ whose magnitude increases with
increasing magnetic field or temperature.%

\begin{figure}
[ptb]
\begin{center}
\includegraphics[
height=6.1145cm,
width=8.3088cm
]%
{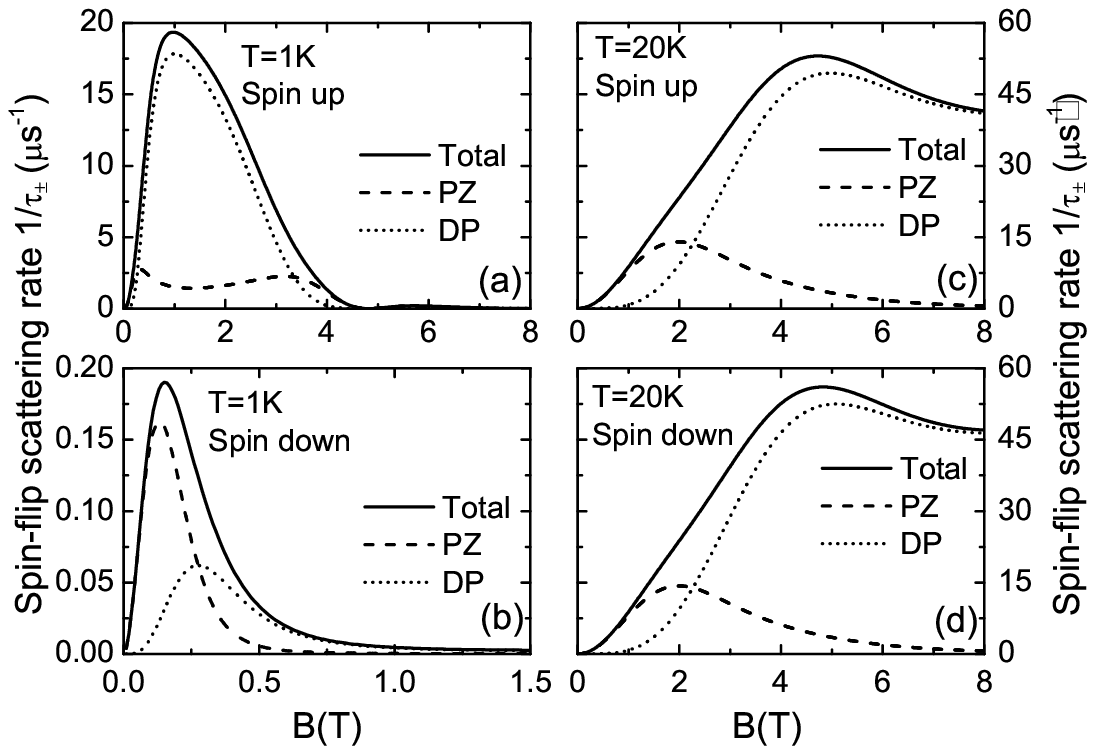}%
\caption{Spin-flip scattering rate of the lowest spin-up level and spin-down
level as a function of magnetic field at $T$=1 K and $T$=20 K, with the same
structure as in Fig. 2. (a) Spin-up level, $T$=1 K; (b) spin-down level, $T$=1
K; (c) spin-up level, $T$=20 K; (d) spin-down level, T=20K. The total
spin-flip scattering rate, the contribution from PZ and DP are denoted by
solid lines, dashed lines, and short-dashed lines, respectively.}%
\end{center}
\end{figure}
The SFR's of the lowest Zeeman doublet $1/\tau_{\pm}$ are shown in Fig. 3 as a
function of magnetic field. The contributions from piezoelectric coupling (PZ)
and deformation potential interaction (DP) are also indicated by the dashed
and short-dashed lines, respectively. First, it is interesting to notice that
$1/\tau_{-}$ in Fig. 3(b) is significantly smaller than $1/\tau_{+}$ in Fig.
3(a), because the transition from $(00-)$ to $(00+)$ needs to absorb a phonon,
but the phonon number is very small at a low temperature $T$=1 K. Second, both
$1/\tau_{+}$ and $1/\tau_{-}$ are suppressed in a strong magnetic field at
$T$=1 K, as shown in Fig. 3(a) and 3(b). The suppression of $1/\tau_{+}$ is
due to the combined effect of small electron spin splitting $\Delta_{0}$ (cf.
Fig. 1(d)) and the vanishing phonon absorption factor $n(\left\vert \Delta
_{0}\right\vert )$, while the suppression of $1/\tau_{-}$ is caused by the
vanishing correlation function $G^{+-}$ (see Fig. 1(a)). We also note that the
PZ contribution dominates at small spin splitting $(\Delta_{0}\lesssim0.3$
meV$),$ while the DP contribution dominates at large spin splitting
$(\Delta_{0}\gtrsim0.3$ meV$)$, which can be clearly seen in Fig. 1(d), Fig.
3(c), 3(d) and all subsequent results. This is a direct result of the
difference between the dependence of the PZ and DP coupling constant
$\alpha_{\nu}(\mathbf{q})$ on the wave vector: $\alpha^{PZ}(\mathbf{q}%
)\propto1/\sqrt{q}$, $\alpha^{DP}(\mathbf{q})\propto\sqrt{q},$ such that the
former (latter) dominates at small (large) phonon energy $\hbar\omega
_{\mathbf{q}}(=\Delta_{0})$. Finally, the electron spin splitting $\Delta_{0}$
in Fig. 2(a) vanishes at $B$=0 T and $B\approx$5 T. Correspondingly, the SFR's
$1/\tau_{\pm}$ vanish in both Fig. 3(a) and Fig. 3(b).

In the high-temperature regime, the spin splitting $\left\vert \Delta
_{0}\right\vert $ increases with increasing magnetic field. Consequently, the
SFR's $1/\tau_{\pm}$ exhibit the same behaviors, as shown in Fig. 3(c) and
3(d). Note, however, the contribution from PZ coupling decreases in a strong
magnetic field. This is caused by the decreasing $S\Gamma$ when the electron
spin splitting $\Delta_{0}$ exceeds $\Delta_{c1}$ (see Fig. 1(d)). Compared
with the low-temperature case, we find that the SFR's increase significantly
at a higher temperature, due to the increasing number of phonons. However, at
very strong magnetic fields, the absence of high-energy phonon and the
reduction of $G^{+-}$, similar to the low-temperature case, reduce $1/\tau
_{+}$ and $1/\tau_{-}$, respectively.

\subsubsection{Intermediate Mn concentration}%

\begin{figure}
[ptb]
\begin{center}
\includegraphics[
height=6.2451cm,
width=8.3066cm
]%
{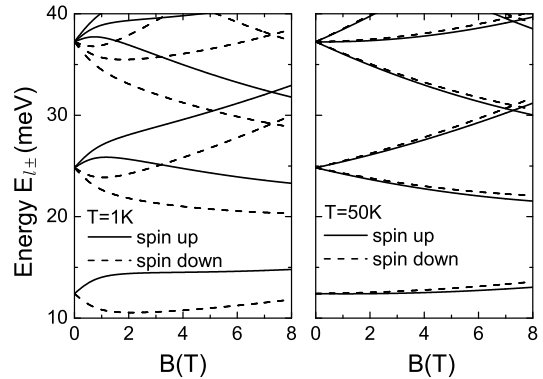}%
\caption{Renormalized spin-dependent electron energy spectra as a function of
magnetic field, at (a) $T$=1 K and (b) $T$=50 K, with fixed $z_{0}$=2 nm,
$d$=16 nm, $x$=0.01. The solid and dashed lines denote the spin-up and
spin-down levels, respectively.}%
\end{center}
\end{figure}
The Mn concentration is increased to $x$=0.01$,$ with a saturated exchange
splitting $(\Delta_{sd})_{sat}$ $\sim$5 meV, large enough to maintain a
positive spin splitting $\Delta_{0}$ in the whole range of the magnetic field
$B$=0$\sim$8 T at low temperature. From the renormalized energy spectra shown
in Fig. 4, we see that $\Delta_{sd}$ ($\Delta_{i}$) dominates at $T$=1 K
($T$=50 K) such that $\Delta_{0}$ is positive (negative) over the whole range
of the magnetic field. In Fig. 5, the SFR's $1/\tau_{\pm}$ are plotted as a
function of magnetic field for $T$=1 K and $T$=50 K, respectively. At $T$=1 K,
when the magnetic field increases, $1/\tau_{+}$ increases to its saturation
value due to the increase and saturation of $\Delta_{sd}$ (and thus
$\Delta_{0}$), while that of the spin-down level is much smaller and shows a
sharp peak at very weak magnetic field $(B\approx$0.04 T$)$ and reduces to
zero quickly, which is caused primarily by the small phonon absorption factor
$n(\left\vert \Delta_{0}\right\vert )$ at low temperature and partly by the
reduction of the correlation function $G^{+-}$ in a strong magnetic field.
\begin{figure}
[ptbptb]
\begin{center}
\includegraphics[
height=6.0739cm,
width=8.3088cm
]%
{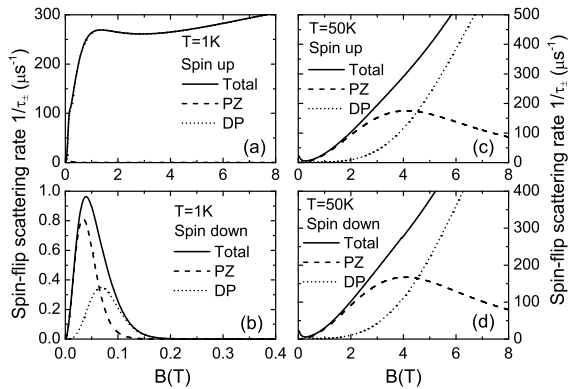}%
\caption{Spin-flip scattering rate of the lowest Zeeman doublet versus the
magnetic field at $T$=1 K and $T$=50 K, with the same structure as in Fig. 4.
(a) Spin-up level, $T$=1 K; (b) spin-down level, $T$=1 K; (c) spin-up level,
$T$=50 K; (d) spin-down level, $T$=50 K. The total spin-flip scattering rate,
the contribution from PZ and DP are denoted by solid, dashed and short-dashed
lines, respectively.}%
\end{center}
\end{figure}
The further increase of $1/\tau_{+}$ with increasing magnetic field in Fig.
5(a) is due to the decrease of the orbital excitation energy (in Eq.
(\ref{Kernel}), the denominator $\Delta_{l_{1}l_{2}}$ consists of two parts,
the orbital excitation energy $\varepsilon_{l_{1}}-\varepsilon_{l_{2}}$ and
the spin splitting $\Delta_{0}$). At $T$=50 K, $1/\tau_{+}$ and $1/\tau_{-}$
both increase with increasing magnetic field, due to the increase of the spin
splitting $\Delta_{0}\ $and the number of phonons. Note in Fig. 5(c) and 5(d),
the zero-field SFR's do not vanish. This can be understood because the
electron can transit to excited orbital levels at $T$=50 K, although the
spin-flip transition to the ground orbital level is prohibited due to
vanishing spin splitting $\Delta_{0}$. Furthermore, we notice the SFR in the
case of intermediate Mn concentration $(x$=0.01$)$ is several times larger
than that for low Mn concentration $(x$=0.002$),$ since $1/\tau_{\pm}$
$\propto x$ through Eq. (\ref{Wsf2_UpDown}) and Eq. (\ref{Wsf2_DownUp}) (note,
however, the effect of the Mn concentration $x$ is also manifested through
changing the electron spin splitting $\Delta_{0}$). Finally, the decrease of
the PZ contribution at large electron spin splitting $\Delta_{0}$, and the
resulted crossing of the PZ and DP contributions at $\Delta_{0}\approx0.3$
meV, as discussed in the previous subsection, is again observed.

\subsubsection{High Mn concentration}%

\begin{figure}
[ptb]
\begin{center}
\includegraphics[
height=6.2699cm,
width=8.3066cm
]%
{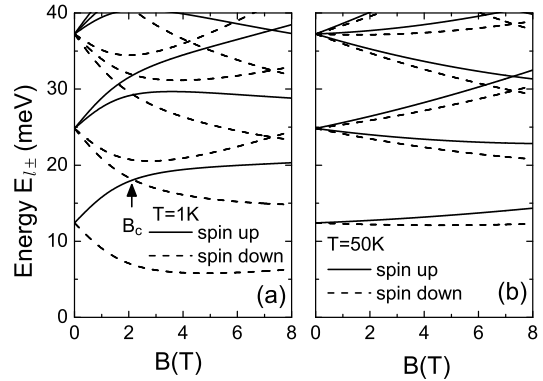}%
\caption{Renormalized spin-dependent electron energy spectra vs. the magnetic
field at (a) $T$=1 K and (b) $T$=50 K, with fixed $z_{0}$=2 nm, $d$=16 nm,
$x$=0.05. Spin-up and spin-down levels are denoted by solid and dashed lines,
respectively.}%
\end{center}
\end{figure}
In this subsection, the Mn concentration is increased further to $x$=0.05$,$
with a saturated exchange splitting $(\Delta_{sd})_{sat}$ $\sim$16 meV, such
that $\Delta_{sd}$ is comparable with the in-plane orbital level separation
($\sim12$ meV). The renormalized energy spectra at $T$=1 K and $T$=50 K are
shown in Fig. 6. We see from the left panel ($T$=1 K) that the spin-up ground
orbital level crosses the spin-down excited orbital level at a critical
magnetic field $B_{c}\sim$2.2 T. At a higher temperature $T$=50 K,
$\Delta_{sd}$ is still the dominant contribution to $\Delta_{0},$ but its
magnitude decreases, such that the energy levels do not cross (see Fig. 6(b)).

The SFR's of the lowest Zeeman doublet are shown in Fig. 7. First we note the
SFR's in panel (a), (c), (d) are of the same order of inverse nanoseconds,
while that in panel (b) is much smaller, due to the absence of high-energy
phonons at low temperature. The most significant feature is the sharp peak
around the critical magnetic field $B_{c}$ in Fig. 7 (a), corresponding to the
crossing of the spin-up ground level with the first excited spin-down level
(see Fig. 6(a)). This is because the level crossing leads to a resonance in
$K_{00,00}$ and, as a result, in $1/\tau_{+}$ (see the discussion in the
beginning of section III).
\begin{figure}
[ptb]
\begin{center}
\includegraphics[
height=6.3668cm,
width=8.3088cm
]%
{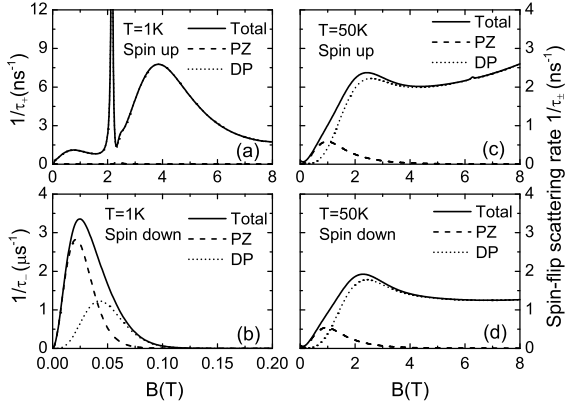}%
\caption{Spin-flip scattering rate of the lowest Zeeman doublet as a function
of magnetic field at $T$=1 K and $T$=50 K, with the same structure as in Fig.
6. (a) Spin-up level, $T$=1 K; (b) spin-down level, $T$=1 K; (c) spin-up
level, $T$=50 K; (d) spin-down level, $T$=50 K. The total spin-flip scattering
rate, the contribution from PZ and DP are denoted by solid lines, dashed
lines, and short-dashed lines, respectively.}%
\end{center}
\end{figure}
Note, however, $1/\tau_{-}$ in Fig. 7(b) doesn't show this resonant behavior,
because the resonance of $K_{00,00}$ is suppressed by the vanishing phonon
absorption factor $n(\left\vert \Delta_{0}\right\vert )$. For $B<B_{c},$ the
transition to the lowest spin-down level gives the dominant contribution to
$1/\tau_{+}$, which reaches its maximum at $B\approx$0.8 T and decreases at
stronger magnetic fields (cf. Fig. 1(d)). For $B>B_{c},$ the electron in the
lowest spin-up level can transit into the first excited spin-down level,
opening up a second spin-flip channel, and it is just the contribution from
this channel that dominates in the $B>B_{c}$ regime. This second contribution
reaches its maximum value at $B\approx$4 T and then decreases, which can also
be interpreted via Fig. 1(d). For the $T$=50 K case, $1/\tau_{+}$ and
$1/\tau_{-}$ both increase with increasing magnetic field, showing a broad
peak at $B\approx$2.5 T. The peak comes from the competing effect of
increasing spin splitting $\Delta_{0}$ (which leads to increasing $1/\tau
_{\pm}$)$\ $against decreasing correlation function $G^{-+},G^{+-}$ and phonon
emission (absorption) factor.

\subsection{Weak lateral confinement}

Now we turn to investigate QD's with weak lateral confinement $d$=40 nm, whose
orbital level separation is comparable with $\Delta_{i}$. In this case, with
small Mn concentration or high temperature, $\Delta_{i}$ makes the main
contribution to the spin splitting. Consequently, the spin splitting
$\Delta_{0}$ is negative, and the lowest spin-down level may cross the excited
spin-up levels in a strong magnetic field. Figure 8 shows the renormalized
electron energy spectra at $T$=1 K and $T$=10 K.
\begin{figure}
[ptb]
\begin{center}
\includegraphics[
height=6.3871cm,
width=8.3088cm
]%
{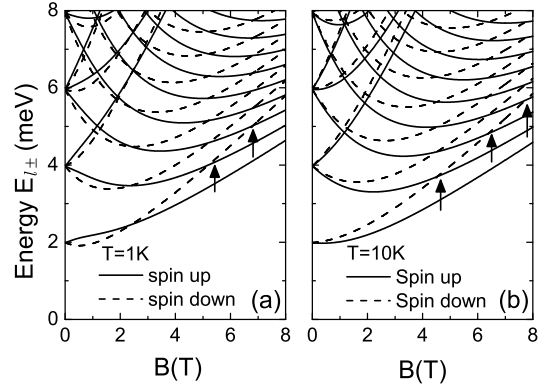}%
\caption{Renormalized spin-dependent electron energy spectra vs. the magnetic
field at (a) $T$=1 K and (b) $T$=10 K, with fixed $z_{0}$=2 nm, $d$=40 nm,
$x$=0.001. Spin-up and spin-down levels are denoted by solid and dashed lines,
respectively.}%
\end{center}
\end{figure}
In panel (a), where the temperature is low, the spin-down ground level first
crosses the spin-up ground level, due to the small exchange splitting
$\Delta_{sd}$ and low temperature, then it sweeps cross the excited spin-up
levels, due to the large $\Delta_{i}$ compared with the orbital excitation
energy in a strong magnetic field. When the temperature increases to $T$=10 K
(see Fig. 8(b)), the exchange splitting is suppressed and $\Delta_{i}$ always
dominates. The lowest spin-down level crosses the excited spin-up levels but
the crossing between the Zeeman split doublet doesn't occur.

Figure 9 shows $1/\tau_{\pm}$ as a function of magnetic field at $T$=1 K and
$T$=10 K, respectively.
\begin{figure}
[ptb]
\begin{center}
\includegraphics[
height=6.2384cm,
width=8.3066cm
]%
{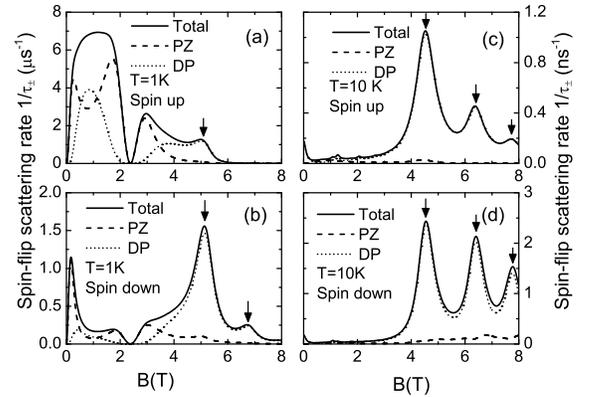}%
\caption{Spin-flip scattering rate of the lowest Zeeman doublet as a function
of magnetic field at $T$=1 K and $T$=10 K, with the same structure as in Fig.
8. (a) Spin-up level, $T$=1 K; (b) spin-down level, $T$=1 K; (c) spin-up
level, $T$=10 K; (d) spin-down level, $T$=10 K. The total spin-flip scattering
rate, the contribution from PZ and DP are denoted by solid lines, dashed
lines, and short-dashed lines, respectively.}%
\end{center}
\end{figure}
In panel (a) and (b), the low-field behaviors of $1/\tau_{\pm}$ resemble those
of strong lateral confinement and low Mn concentration (see Fig. 3(a) and
3(b)). At higher temperature, $1/\tau_{\pm}$ exhibit many peaks at higher
magnetic fields (indicated by the arrows), which are caused by the
aforementioned level crossings. However, in very strong magnetic fields, the
peaks are suppressed by the phonon absorption factor (for $1/\tau_{+}$) and
the correlation function $G^{+-}$ (for $1/\tau_{-}$). At a higher temperature
$T$=10 K, the resonances of the kernel $K_{00,00}$ are less suppressed and
more pronounced peaks arises in $1/\tau_{\pm}$, leading to short spin
lifetimes of the order of nanoseconds, compared with the microsecond scale in
the $T$=1 K case.

\subsection{Temperature effect}

In the above, we have observed that the temperature plays an important role in
determining the SFR through changing the electron spin splitting $\Delta_{0}$,
the correlation function $G^{+-}$, $G^{-+}$ and the phonon emission
(absorption) factor. Taking a small QD ($z_{0}$=2 nm, $d$=16 nm) for example,
we plot the renormalized energy spectrum and $1/\tau_{\pm}$ as a function of
temperature at $B$=4 T in Fig. 10.
\begin{figure}
[ptb]
\begin{center}
\includegraphics[
height=6.4186cm,
width=8.3088cm
]%
{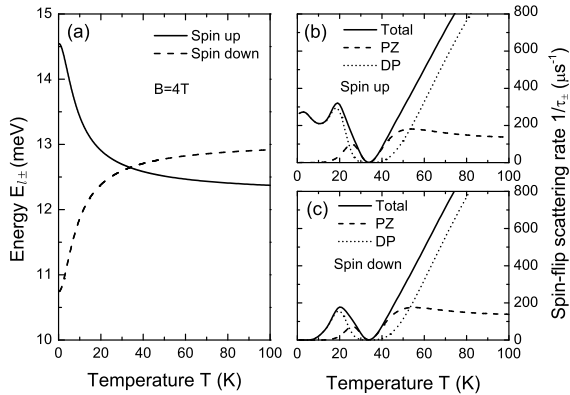}%
\caption{(a) Renormalized energy of the lowest spin-up (solid line) and
spin-down (dashed line) levels vs. the temperature. (b) and (c) show the the
spin-flip scattering rate of the spin-up and spin-down levels vs. the
temperature, respectively. The total spin-flip scattering rate, the PZ and DP
contribution are denoted by solid, dashed and short-dashed lines,
respectively. Here we take $B$=4 T, z$_{0}$=2 nm$,$ $d$=16 nm and $x$=0.01 in
the numerical calculation.}%
\end{center}
\end{figure}
It is interesting to notice that the Zeeman split doublet crosses each other
at an elevated temperature (see Fig. 10(a)), due to the reduction of the
exchange splitting $\Delta_{sd}$. From Fig. 1(d), we see that the quantity
$S\Gamma$ decreases with decreasing $\Delta_{0}$ for small $\Delta_{0}$. This
effect, together with the phonon emission factor, which shows a sharp peak at
$\Delta_{0}$=0$,$ results in the non-monotonic temperature dependence of
$1/\tau_{+}$ shown in Fig. 10(b). In Fig. 10(c), the low-temperature SFR for
the spin-down level $1/\tau_{-}$ vanishes due to the absence of high-energy
phonons and the vanishing correlation function $G^{+-}$. Note at $T\approx$34
K, the SFR of both spin-up and spin-down levels vanishes, due to the vanishing
spin splitting $\Delta_{0}$.

\subsection{Dependence of the SFR on the confinement}

Both the vertical and lateral confinement of the QD can affect the spin
relaxation significantly through varying the electron wave function. The
effect of vertical confinement on $1/\tau_{\pm}$ comes from the form factor
$Z(q)$ (see Eq. (\ref{ConfineFactor})) and the overlap integral (see Eq.
(\ref{OverlapIntegral})). It can be seen from Fig. 1(c) that the quantity
$S\Gamma$ is roughly proportional to $1/z_{0},$ such that the SFR's should
also show the same behavior, which can be seen in Fig. 11(a) and 11(b). Note
here the largest $z_{0}$ (20nm) still sustains a vertical orbital level
separation of $\sim30$ meV, such that the influence of higher subbands on the
spin relaxation is negligibly small. The approximate relationship $1/\tau
_{\pm}\propto1/z_{0}$ comes from two factors. First, the \textit{s-d} exchange
scattering amplitude with one Mn ion is proportional to $1/z_{0}$ which, when
squared (in the Fermi golden rule) and averaged over all the Mn sites, leads
to the $1/z_{0}$ dependence of the spin-flip scattering rate to a given final
state. Second, the spin-flip channel (i.e., the number of final states)
doesn't increase provided $z_{0}$ is small enough such that only the lowest
bound state is relevant. We notice that G. Bastard et al. performed a
theoretical calculation of the SFR of subbands in DMS quantum wells, and
similar dependence of the SFR on the well width is predicted.\cite{Bastard}%

\begin{figure}
[ptb]
\begin{center}
\includegraphics[
height=6.1528cm,
width=8.3088cm
]%
{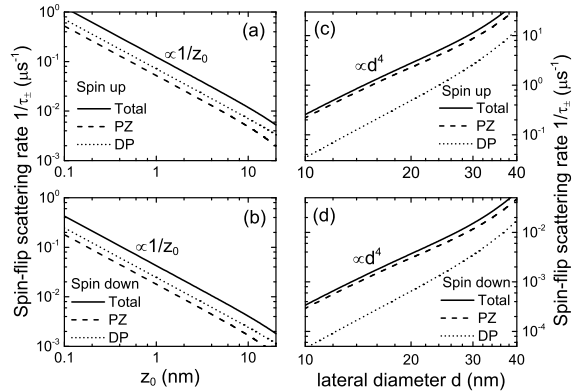}%
\caption{(a) and (b) show 1/$\tau_{\pm}$ vs. the vertical confining length
$z_{0}$ at fixed $B$=1 T, $T$=1 K, $d$=16 nm, $x$=0.001. (c) and (d) show
1/$\tau_{\pm}$ vs. the lateral diameter $d$ at fixed $B$=4 T, $T$=1 K,
$x$=0.002. The total spin-flip scattering rate, the PZ and DP contributions
are denoted by solid, dashed and short-dashed lines, respectively.}%
\end{center}
\end{figure}
The effect of the lateral confinement strength, characterized by the lateral
diameter $d$, on the spin relaxation is shown in Fig. 11(c) and 11(d). The
$d^{4}$ dependence of $1/\tau_{\pm}$ can be appreciated as follows. From the
Appendix and the definition $d=2\sqrt{\hbar/(m^{\ast}\omega_{0})}$, we see the
orbital level separation $\delta$ is roughly proportional to $1/d^{2}$, while
$1/\tau_{\pm}$ is inversely proportional to $\delta^{2}$ (see Eq.
(\ref{Kernel})) when the spin splitting is small, so we expect that
$1/\tau_{\pm}$ should be approximately proportional to $d^{4}$, although the
precise dependence of $1/\tau_{\pm}$ on $d$ is also affected by the
phonon-induced transition rate $\Gamma$ (cf. Eq. (\ref{PZPhonon}),
(\ref{DPPhonon})).

The dependence of the SFR on the QD size $1/\tau_{\pm}\propto z_{0}^{-1}d^{4}$
caused by phonon-mediated \textit{s-d} exchange scattering is quite different
from those caused by other spin relaxation mechanisms in nonmagnetic
semiconductor QD's.\cite{Woods,Smimov,MWWu} We note that this relationship can
be deduced from the work by Nazarov,\cite{Nazarov2} where the spin relaxation
in nonmagnetic semiconductor QD's is considered, but the magnitude of the SFR
in our results is several orders of magnitude higher.

\section{Conclusions}

Based on second-order perturbation theory, we have investigated the SFR caused
by the phonon-mediated \textit{s-d} exchange interaction of the lowest Zeeman
split doublet in II-VI DMS QD's as a function of magnetic field, as well as
the dependence of the SFR on the Mn concentration, dot size and temperature.
We found the SFR increases with increasing Mn concentration and electron spin
splitting $\Delta_{0}$. Increasing the lateral dot size leads to enhanced SFR
while increasing the vertical dot size reduces the SFR for a small QD. The
dependence of the SFR on the magnetic field and temperature shows
non-monotonic behaviors, due to the competing effect between the electron spin
splitting, the phonon emission (absorption) factor and the correlation
function of the Mn ions. It is interesting to notice that the spin relaxation
of both spin-up and spin-down electrons is suppressed in the case of strong
magnetic field and low Mn concentration at low temperature.

\begin{acknowledgments}
This work was supported by the NSFC No. 60376016, 863 project No. 2002AA31,
and the special fund for Major State Basic Research Project No. G001CB3095 of China.
\end{acknowledgments}

\appendix*

\section{EXPRESSIONS OF $S$ AND $\Gamma$}

The orbital part of $\bar{H}_{0}$ gives the Fock-Darwin states%
\begin{align}
\phi_{nm}(\rho,\varphi)  &  =\frac{1}{\sqrt{2\pi}}e^{im\varphi}\cdot
R_{nm}(\rho)\ \ (n,\left\vert m\right\vert =0,1,2,\cdots
),\label{OrbitalEnergy}\\
R_{nm}(\rho)  &  =\frac{\sqrt{2}}{l_{0}}\sqrt{\frac{n!}{(n+\left\vert
m\right\vert )!}}(\frac{\rho}{l_{0}})^{\left\vert m\right\vert }\exp
(-\frac{\rho^{2}}{2l_{0}^{2}})L_{n}^{\left\vert m\right\vert }(\frac{\rho^{2}%
}{l_{0}^{2}}),\nonumber
\end{align}
with corresponding orbital energy $\varepsilon_{nm}=(2n+\left\vert
m\right\vert +1)\hbar\omega+m\hbar\omega_{c}/2,$ where $l_{0}=\sqrt
{\hbar/(m^{\ast}\omega)},$ $L_{n}^{m}(x)\ $is the generalized Laguerre
polynomial. For convenience, we also introduce $n_{+}=n+(\left\vert
m\right\vert +m)/2,$ $n_{-}=n+(\left\vert m\right\vert -m)/2.$

The dimensionless overlap integral is \begin{widetext}
\begin{align}
S_{l_{1}l_{2}l_{3}l_{4}}  &  =\delta_{m_{1}+m_{2},m_{3}+m_{4}}\sqrt
{\frac{n_{1}!n_{2}!n_{3}!n_{4}!}{(n_{1}+\left\vert m_{1}\right\vert
)!(n_{2}+\left\vert m_{2}\right\vert )!(n_{3}+\left\vert m_{3}\right\vert
)!(n_{4}+\left\vert m_{4}\right\vert )!}}\frac{\Omega\xi}{\pi z_{0}l_{0}^{2}%
}\label{OverlapIntegral}\\
&
{\displaystyle\int_{0}^{\infty}}
e^{-2x}(\sqrt{x})^{\left\vert m_{1}\right\vert +\left\vert m_{2}\right\vert
+\left\vert m_{3}\right\vert +\left\vert m_{4}\right\vert }L_{n_{1}%
}^{\left\vert m_{1}\right\vert }(x)L_{n_{2}}^{\left\vert m_{2}\right\vert
}(x)L_{n_{3}}^{\left\vert m_{3}\right\vert }(x)L_{n_{4}}^{\left\vert
m_{4}\right\vert }(x)dx,\nonumber
\end{align}
\end{widetext}
where $\xi=z_{0}\int dz\left\vert \chi(z)\right\vert ^{4}=3/2$. The phonon
transition rate due to piezoelectric coupling to the acoustic phonon
is\begin{widetext}
\begin{align}
\Gamma_{l_{1}l_{1}^{\prime}l_{2}l_{2}^{\prime}}^{PZ}  &  =\delta_{m_{1}%
-m_{2},m_{1}^{\prime}-m_{2}^{\prime}}\sqrt{\frac{(n_{1+,<})!(n_{1-,<}%
)!(n_{2+,<})!(n_{2-,<})!}{(n_{1+,>})!(n_{1-,>})!(n_{2+,>})!(n_{2-,>})!}%
}(-1)^{\left\vert n_{2+}-n_{2+}^{\prime}\right\vert +\left\vert n_{2-}%
-n_{2-}^{\prime}\right\vert +\frac{N}{2}}\dfrac{(eh_{14})^{2}}{4\pi\hbar\rho
}\label{PZPhonon}\\
&
{\displaystyle\sum\limits_{\nu}}
(\dfrac{l_{0}q_{\nu}}{2})^{N}\dfrac{q_{\nu}}{C_{\nu}^{2}}%
{\displaystyle\int_{0}^{\pi}}
A_{\nu}(\theta)\left\vert Z(q_{\nu}\cos\theta)\right\vert ^{2}e^{-\frac{1}%
{2}(l_{0}q_{\nu}\sin\theta)^{2}}%
\mathcal{F}%
(\dfrac{l_{0}^{2}q_{\nu}^{2}\sin^{2}\theta}{4})(\sin\theta)^{N+1}%
d\theta.\nonumber
\end{align}
\end{widetext}
The contribution from the deformation potential interaction
is\begin{widetext}
\begin{align}
\Gamma_{l_{1}l_{1}^{\prime}l_{2}l_{2}^{\prime}}^{DP}  &  =\delta_{m_{1}%
-m_{2},m_{1}^{\prime}-m_{2}^{\prime}}\sqrt{\frac{(n_{1+,<})!(n_{1-,<}%
)!(n_{2+,<})!(n_{2-,<})!}{(n_{1+,>})!(n_{1-,>})!(n_{2+,>})!(n_{2-,>})!}%
}(-1)^{\left\vert n_{2+}-n_{2+}^{\prime}\right\vert +\left\vert n_{2-}%
-n_{2-}^{\prime}\right\vert +\frac{N}{2}}\frac{\Xi_{d}^{2}q_{l}^{3}}{4\pi
\hbar\rho c_{l}^{2}}(\frac{q_{l}l_{0}}{2})^{N}\label{DPPhonon}\\
&  \int_{0}^{\pi}(\sin\theta)^{N+1}\left\vert Z(q_{l}\cos\theta)\right\vert
^{2}\exp(-\frac{q_{l}^{2}l_{0}^{2}\sin^{2}\theta}{2})%
\mathcal{F}%
(\frac{q_{l}^{2}l_{0}^{2}\sin^{2}\theta}{4})d\theta.\nonumber
\end{align}
\end{widetext}
In the above, we have used the short notation $l$ to denote the quantum number
$(n,m)$ and $n_{j,<}\ (n_{j,>})$ to denote $\min\left\{  n_{j},n_{j}^{\prime
}\right\}  $ ($\max\left\{  n_{j},n_{j}^{\prime}\right\}  $), e.g., $l_{1}$
stands for $(n_{1},m_{1})$ and $n_{1+,>}$ stands for $\max\left\{
n_{1+},n_{1+}^{\prime}\right\}  $. Other quantities are $N=\left\vert
n_{1+}-n_{1+}^{\prime}\right\vert +\left\vert n_{1-}-n_{1-}^{\prime
}\right\vert +\left\vert n_{2+}-n_{2+}^{\prime}\right\vert +\left\vert
n_{2-}-n_{2-}^{\prime}\right\vert ,$ $q_{\nu}=\left\vert \Delta_{ll^{\prime}%
}\right\vert /(\hbar C_{\nu}),$ $C_{\nu}$ ($\nu=l,t$) is the longitudinal or
transverse sound velocity$,$ $\theta$ is the polar angle of the phonon vector
$\mathbf{q},$ $A_{\nu}(\theta)$ is the anisotropy function of the
piezoelectric interaction, with the dependence on the the azimuth angle
$\varphi$ averaged out,%
\begin{equation}
Z(q)=%
{\displaystyle\int}
e^{iqz}\left\vert \chi(z)\right\vert ^{2}dz=\frac{4\pi^{2}i(e^{iqz_{0}}%
-1)}{qz_{0}\left[  (qz_{0})^{2}-(2\pi)^{2}\right]  } \label{ConfineFactor}%
\end{equation}
is the form factor, $%
\mathcal{F}%
(x)=L_{n_{1+,<}}^{\left\vert n_{1+}-n_{1+}^{\prime}\right\vert }%
(x)L_{n_{1-,<}}^{\left\vert n_{1-}-n_{1-}^{\prime}\right\vert }(x)L_{n_{2+,<}%
}^{\left\vert n_{2+}-n_{2+}^{\prime}\right\vert }(x)L_{n_{2-,<}}^{\left\vert
n_{2-}-n_{2-}^{\prime}\right\vert }(x),$ and $\Xi_{d}$ is the deformation
potential constant.

For the spin-flip transitions between the lowest Zeeman doublet $\left\vert
00+\right\rangle $ and $\left\vert 00-\right\rangle ,$ the following overlap
integral and phonon transition rates are used:\begin{widetext}
\begin{equation}
S_{00,n_{1}m_{1},00,n_{2}m_{2}}=\delta_{m_{1},m_{2}}\dfrac{(n_{1}%
+n_{2}+\left\vert m_{1}\right\vert )!}{2^{n_{1}+n_{2}+\left\vert
m_{1}\right\vert }\sqrt{n_{1}!n_{2}!(n_{1}+\left\vert m_{1}\right\vert
)!(n_{2}+\left\vert m_{2}\right\vert )!}}\frac{\Omega\xi}{2\pi z_{0}l_{0}^{2}%
}, \label{OverlapGround}%
\end{equation}%
\begin{align}
\Gamma_{00,n_{1}m_{1},00,n_{2}m_{2}}^{PZ}  &  =\delta_{m_{1},m_{2}}%
\dfrac{(-1)^{n_{1}+n_{2}}}{\sqrt{n_{1}!(n_{1}+\left\vert m_{1}\right\vert
)!n_{2}!(n_{2}+\left\vert m_{2}\right\vert )!}}\dfrac{(eh_{14})^{2}}{4\pi
\hbar\rho}%
{\displaystyle\sum\limits_{\nu}}
(\dfrac{l_{0}q_{\nu}}{2})^{2(n_{1}+n_{2}+\left\vert m_{1}\right\vert )}%
\dfrac{q_{\nu}}{C_{\nu}^{2}}\label{PZPhononGround}\\
&
{\displaystyle\int_{0}^{\pi}}
A_{\nu}(\theta)\left\vert Z(q_{\nu}\cos\theta)\right\vert ^{2}\exp
(-\frac{l_{0}^{2}q_{\nu}^{2}\sin^{2}\theta}{2})%
\mathcal{F}%
(\dfrac{l_{0}^{2}q_{\nu}^{2}\sin^{2}\theta}{4})(\sin\theta)^{2(n_{1}%
+n_{2}+\left\vert m_{1}\right\vert )+1}d\theta,\nonumber
\end{align}%
\begin{align}
\Gamma_{00,n_{1}m_{1},00,n_{2}m_{2}}^{DP}  &  =\delta_{m_{1},m_{2}}%
\frac{(-1)^{n_{1}+n_{2}}}{\sqrt{n_{1}!n_{2}!(n_{1}+\left\vert m_{1}\right\vert
)!(n_{2}+\left\vert m_{2}\right\vert )!}}\frac{\Xi_{d}^{2}q_{l}^{3}}{4\pi
\hbar\rho c_{l}^{2}}(\frac{q_{l}l_{0}}{2})^{2(n_{1}+n_{2}+\left\vert
m_{1}\right\vert )}\label{DPPhononGround}\\
&  \int_{0}^{\pi}(\sin\theta)^{2(n_{1}+n_{2}+\left\vert m_{1}\right\vert
)+1}\left\vert Z(q_{l}\cos\theta)\right\vert ^{2}\exp(-\frac{q_{l}^{2}%
l_{0}^{2}\sin^{2}\theta}{2})%
\mathcal{F}%
(\frac{q_{l}^{2}l_{0}^{2}\sin^{2}\theta}{4})d\theta,\nonumber
\end{align}
\end{widetext}
and $S_{n_{1}m_{1},n_{2}m_{2},00,00}$, $\Gamma_{00,n_{1}m_{1},n_{2}m_{2}%
,00}^{PZ}$, $\Gamma_{00,n_{1}m_{1},n_{2}m_{2},00}^{DP}$ can be obtained from
$S_{00,n_{1}m_{1},00,n_{2}m_{2}},$ $\Gamma_{00,n_{1}m_{1},00,n_{2}m_{2}}%
^{PZ},$ $\Gamma_{00,n_{1}m_{1},00,n_{2}m_{2}}^{DP}$, respectively, by
replacing $\delta_{m_{1},m_{2}}$ with $\delta_{m_{1},-m_{2}}$.

\end{document}